# FDTD Simulation of O-X Mode Conversion Process in Non-uniform Magnetized Plasma

Chenxu Wang[1], Ryota Usui[2], Hiroaki Nakamura[1,2], Hideki Kawaguchi[3], Kubo Shin[4]

[1]National Institute for Fusion Science, 322-6 Oroshi, Toki, 509-5292 Japan
[2]Nagoya University, Furo-cho, Chikusa-ku, Nagoya 464-8603, Japan
[3]Muroran Institute of Technology, Muroran, Hokkaido 050-0071, Japan
[4]Chubu University, Kasugai, Aichi 487-8501. Japan

e-mail: naka-lab@nifs.ac.jp

Electron Bernstein Waves (EBWs) are electrostatic waves than can propagate in overdense plasmas without density cutoff, making them suitable for high density plasma heating. Since EBWs cannot be directly launched from vacuum, mode conversion processes such as O-X-B conversion are required. In this study, the O-X mode conversion process is investigated using the finite difference time-domain (FDTD) method in a magnetized plasma with non-uniform density. In particular, the dependence of mode conversion characteristics on the incident angle of the injected wave is studied. The results show that an optimal incident angle exists at which the wave propagates without significant attenuation and strong electric field enhancement is observed near the upper hybrid resonance (UHR) layer. When the incident angle deviates from this optimal angle, an evanescent region appears, resulting in attenuation of the wave. These results demonstrate that angular optimization is essential for efficient wave propagation toward the UHR region and EBW excitation.
Key words: EBWs, plasma heating, mode conversion, FDTD, non-uniform density

1. INTRODUCTION

Efficient heating of overdense plasmas remains a critical issue in magnetic confinement fusion research. Conventional electron cyclotron resonance heating (ECRH) is limited by the cutoff density, beyond which electromagnetic waves cannot penetrate into the plasma core [1, 2]. Therefore, the development of alternative heating schemes capable of delivering power into overdense plasmas is of significant importance. Electron Bernstein Waves (EBWs), which are electrostatic waves in magnetized plasmas, have attracted considerable attention because they do not suffer from density cutoff and can be strongly absorbed at electron cyclotron harmonics [3-5]. However, since EBWs cannot propagate in vacuum, mode conversion processes such as O-X-B conversion are required to achieve a highly efficiency plasma heating [6, 7]. Recent studies have explored alternative wave structures for improving plasma heating performance. In particular, electromagnetic waves carrying orbital angular momentum (OAM), such as millimeter-wave vortex fields [8, 9], have been investigated as a potential candidate for efficient energy delivery into plasmas, providing a possible route toward enhanced plasma heating [10]. In our previous work, we investigated the propagation characteristics of millimeter-wave vortex field in a magnetized plasma with a uniform density profile using the finite-difference time-domain (FDTD) method. It was demonstrated that the vortex wave can penetrate into plasma regions where plane waves cannot propagate [11, 12]. However, in realistic fusion plasmas, the density is inherently non-uniform, and wave propagation is strongly influenced by spatial gradients. Under such conditions, mode conversion phenomena, become essential for describing wave penetration and energy transport. Therefore, it is necessary to extend previous studies to inhomogeneous plasma configurations and to clarify the process of mode conversion in wave propagation. In this study, we investigate the O-X mode conversion process in a non-uniform magnetized plasma using the FDTD method. In particular, we discuss the dependence of wave propagation characteristics on the incident angle.

## 2. THEORETICAL OF O-X MODE CONVERSION IN PLASMA

Several mode conversion schemes have been proposed for electron Bernstein wave (EBW) excitation, including the O–X–B, X–B, and SX–B processes. Among these, the O–X–B scheme is widely adopted due to its flexibility in experimental configurations and its strong dependence on the incident angle. In the present study, we focus on the O–X conversion stage, which determines whether the injected electromagnetic wave can penetrate into the plasma and reach the upper hybrid resonance (UHR) region. This stage plays a fundamental role in EBW excitation and is therefore essential for understanding wave propagation in non-uniform plasmas.

### 2.1 DISPERSION CHARACTERISTICS OF O- AND X-MODES

The Propagation of electromagnetic waves in a magnetized plasma is governed by anisotropic dispersion relations, which give rise to distinct wave modes depending on the polarization and propagation direction relative to the magnetic field. For perpendicular propagation, there exist two propagation modes in magnetized plasma, that is ordinary (O) mode and extraordinary (X) mode. The O-mode is characterized by a cutoff at $\omega = \omega_{pe}$, beyond which wave propagation is not allowed. In contrast, the X-mode exhibits a more complex dispersion structure, including both cutoff and resonance layers, such as the right-hand cutoff ($\omega_R$), left-hand cutoff ($\omega_L$), and the UHR, defined by $\omega^2 = \omega_{pe}^2 + \omega_{ce}^2$. Fig. 1 shows a representative dispersion relation of the O-mode and X-mode. As illustrated, the O-mode branch terminates at the cutoff layer, while the X-mode branch extends toward higher plasma density and connects to the UHR region. Near the UHR layer, the perpendicular refractive index increases rapidly, indicating strong wave plasma interaction. These dispersion characteristics determine whether an externally launched wave can access the overdense plasma region and form the basis for the O-X mode conversion process discussed in the following sections.

### 2.2 O-X MODE CONVERSION

When an electromagnetic wave is injected obliquely into a plasma with a density gradient, the parallel refractive index, $n_\parallel = sin\theta$, remains conserved. Under this condition, the wave evolves continuously along the density gradient can be described as a coupled mode system. As the wave approaches the cutoff layer, the dispersion branches of the O-like and X-like modes become close to each other, allowing mode coupling. This process leads to partial conversion of the incident O-mode into an X-mode, which can propagate further into the plasma. Fig. 2 shows the perpendicular refractive index $n_\perp^2$ as a function of electron density for different incident angles relative to the optimal value. The vertical line indicates the location of the UHR. At the optimal incident angle, the perpendicular refractive index $n_\perp^2$ remains positive throughout the propagation region, indicating that no evanescent region is formed. Under this condition, the wave can undergo continuous O-X mode conversion without significant attenuation. In contrast, when the incident angle deviates from the optimal value, a region where $n_\perp^2 < 0$ appears between the cutoff layer and the UHR region, as clearly seen in Fig.2. In this region, the perpendicular wave number becomes imaginary, leading to exponential attenuation of the wave amplitude. The width of this evanescent region increases as the deviation from the optimal angle becomes larger, resulting in stronger attenuation and reduced wave penetration. Consequently, the amount of wave energy that can reach the UHR region is significantly reduced.

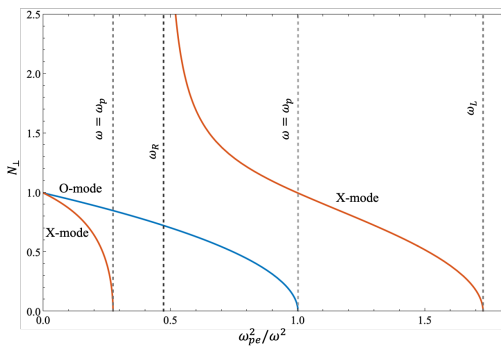

Fig.1 Dispersion relation of the O-mode and X-mode in a magnetized plasma. The perpendicular refractive index is plotted as a function of the normalized plasma density.

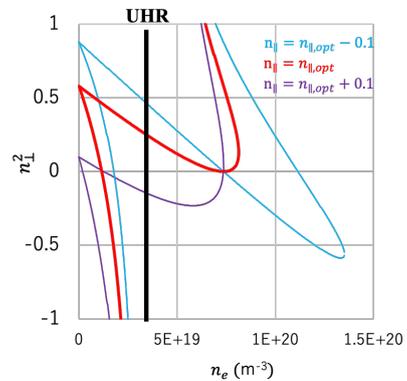

Fig.2 Perpendicular refractive index $n_\perp^2$ as a function of electron density $n_e$ for different values of the parallel refractive index $n_\parallel$ relative to the optimal condition. The vertical line indicates the position of UHR.

## 3. NUMERICAL MODEL AND FDTD SIMULATION

The propagation of electromagnetic waves in the magnetized plasma is simulated using the finite-difference time-domain (FDTD) method. Maxwell's equations are solved in the time domain using Yee grids, in which the electric and magnetic fields are staggered in both space and time. The governing equations are given by,

$$\nabla \times \boldsymbol{E} = -\frac{\partial \boldsymbol{B}}{\partial t}, \tag{1}$$

$$\nabla \times \boldsymbol{B} = \mu_0 \left( \boldsymbol{J} + \varepsilon_0 \frac{\partial \boldsymbol{E}}{\partial t} \right), \tag{2}$$

where J represents the plasma current density.
We adopt the following Drude-Lorentz macro model for the full simulation of plasma particles by using electron displacement density vector $P$ and current density vector $\boldsymbol{J} = d\boldsymbol{P}/dt$,

$$\frac{d\boldsymbol{J}}{dt} + \gamma \boldsymbol{J} + \omega_0^2 \boldsymbol{P} = \varepsilon_0 \omega_p^2 \left( \boldsymbol{E} + \frac{1}{n_e q_e} \boldsymbol{J} \times \boldsymbol{B}_0 \right) \tag{3}$$

where $\gamma$, $n_e$, $q_e$ and $\boldsymbol{B}_0$ are the dumping coefficient, the electron density, elementary charge and externally applied magnetic field, respectively. Then the FDTD analysis of the millimeter-wave vortex in magnetized plasma in 3D grid space for $\boldsymbol{E}$, $\boldsymbol{H}$, $\boldsymbol{P}$ and $\boldsymbol{J}$ is as follows,

$$\boldsymbol{E}^{n+1} = \frac{\frac{\varepsilon_0}{\Delta t} - \frac{\sigma}{2}}{\frac{\varepsilon_0}{\Delta t} + \frac{\sigma}{2}} \boldsymbol{E}^n + \frac{1}{\frac{\varepsilon_0}{\Delta t} + \frac{\sigma}{2}} \nabla \times \boldsymbol{H}^{n+\frac{1}{2}} - \frac{1}{\frac{\varepsilon_0}{\Delta t} + \frac{\sigma}{2}} \boldsymbol{J}^{n+\frac{1}{2}}, \tag{4}$$

$$\boldsymbol{H}^{n+\frac{1}{2}} = \boldsymbol{H}^{n-\frac{1}{2}} - \frac{\Delta t}{\mu_0} \nabla \times \boldsymbol{E}^n, \tag{5}$$

$$\boldsymbol{P}^{n+1} = \Delta t \boldsymbol{J}^{n+\frac{1}{2}} + \boldsymbol{P}^n, \tag{6}$$

$$\frac{\boldsymbol{J}^{n+\frac{1}{2}} - \boldsymbol{J}^{n-\frac{1}{2}}}{\Delta t} + \gamma \frac{\boldsymbol{J}^{n+\frac{1}{2}} + \boldsymbol{J}^{n-\frac{1}{2}}}{2} + \omega_0^2 \boldsymbol{P}^n = \varepsilon_0 \omega_p^2 \boldsymbol{E}^n + \frac{\varepsilon_0 \omega_p^2}{q_e n_e} \frac{\boldsymbol{J}^{n+\frac{1}{2}} + \boldsymbol{J}^{n-\frac{1}{2}}}{2} \times \boldsymbol{B}_0, \tag{7}$$

where $\Delta t$ is unit time step, $\boldsymbol{E}$ and $\boldsymbol{P}$ are assigned to integer time step, $\boldsymbol{H}$ and $\boldsymbol{J}$ are assigned to half integer time step, that is, $\boldsymbol{E}$ and $\boldsymbol{P}$ or $\boldsymbol{H}$ and $\boldsymbol{J}$ are calculated simultaneously. In this model, kinetic effects such as finite larmor radius corrections are not included. Therefore, the present study focuses on the O-X mode conversion process, which is governed primarily by cold plasma dispersion properties.

## 4. NUMERICAL EXAMPLES

The numerical model used in this study is illustrated in Fig.3. An O-mode millimeter-wave is injected obliquely from a waveguide into a magnetize plasma with a non-uniform density profile. The plasma density increases along the radial direction (Fig.4) and is described by

$$n_e = \frac{2\omega^2 \epsilon_0 m_e}{q^2} \times \left(1 - \exp\left(-\frac{r}{a}\right)\right) \tag{8}$$

with a characteristic scale length $a = 0.01\ m$. The external magnetic field is set to $B_0 = 2.0\text{T}$, and the wave frequency is $f = 77$ GHz. The optimal incident angle predicted from the theoretical condition is $\theta_{opt} = 40.45°$. Simulations are performed for both the optimal and non-optimal cases to investigate the dependence of wave propagation on the incident angle. Fig. 5 shows the electric field distribution for the optimal incident angle $\theta = 40.45°$. The injected O-mode wave propagates toward the plasma boundary and undergoes mode conversion near the cutoff region. After the conversion, the wave continues to propagate into the higher density region as an X-mode without significant attenuation. A clear enhancement of the electric field amplitude is observed near the UHR region. This behavior is consistent with the theoretical prediction that the group velocity decreases near the UHR, leading to localization of wave energy. Furthermore, as indicated by the corresponding dispersion relation, no region with $n_\perp^2 < 0$ exists between the conversion point and the UHR. This confirms that the wave does not encounter an evanescent barrier and can propagate continuously through the plasma.

Fig. 6 shows the electric field distribution for a non-optimal incident angle $\theta = 30°$. In this case, the injected wave undergoes strong attenuation after passing the cutoff region. The electric field amplitude decreases significantly before reaching the UHR layer, and only a weak field enhancement is observed. The corresponding dispersion relation shows the presence of a region where $n_\perp^2 < 0$, indicating an evanescent layer. As a result, the perpendicular wave number becomes imaginary, leading to exponential decay of the wave amplitude. This demonstrates that the deviation from the optimal incident angle leads to the formation of an evanescent region, which suppresses wave penetration into the plasma.

## 5. CONCLUSION

In this study, the O-X mode conversion process in a magnetized plasma with non-uniform density has been investigated using the FDTD method. In particular, the dependence of wave propagation characteristics on the incident angle was investigated. The numerical results are in good agreement with the theoretical predictions. At the optimal incident angle, the injected O-mode wave is efficiently converted into the X-mode and propagates toward the UHR region without significant attenuation. Strong electric field enhancement is observed near the UHR, indicating effective wave energy localization. In contrast, for non-optimal incident angles, an evanescent region appears, leading to exponential attenuation of the wave and significantly reduced field amplitude near the UHR. These results demonstrate that the incident angle affect the accessibility of wave propagation and the effectiveness of O-X mode conversion in non-uniform plasmas. And the results provide important insights into the conditions required for efficient EBW excitation. Future work will include evaluation of power flow and conversion efficiency, as well as extension to full O-X-B conversion processes incorporating kinetic effects.

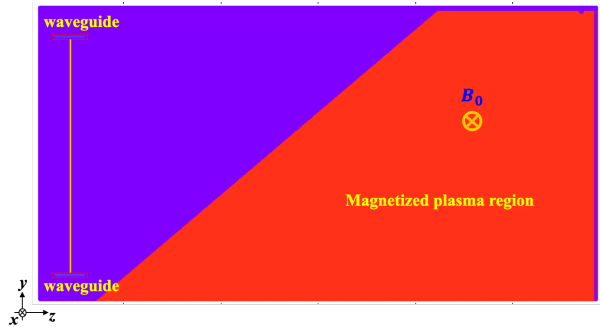

Fig.3 FDTD simulation of numerical model (y-z vertical cross-section) for the propagation of millimeter-wave in magnetized plasma

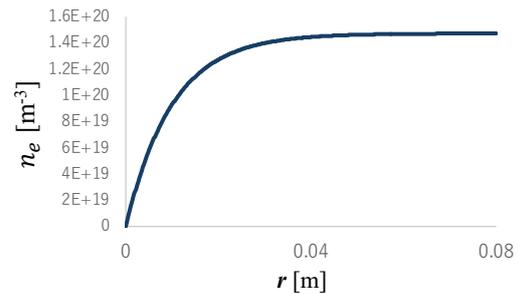

Fig. 4 The distribution of plasma density along the radial direction

## ACKNOWLEDGEMENTS

The computation was performed using Research Center for Computational Science, Okazaki, Japan (Project: 25-IMS-C100) and Plasma Simulator of NIFS. The research was supported by KAKENHI (Nos. 21H04456, 22K03572, 23K03362, 23K11190, 24K00613, 25H01640), by the NINS program of Promoting Research by Networking among Institutions (01422301) by the NIFS Collaborative Research Programs (NIFS24KIG002, NIFS25KIST062, NIFS25KIIT019).

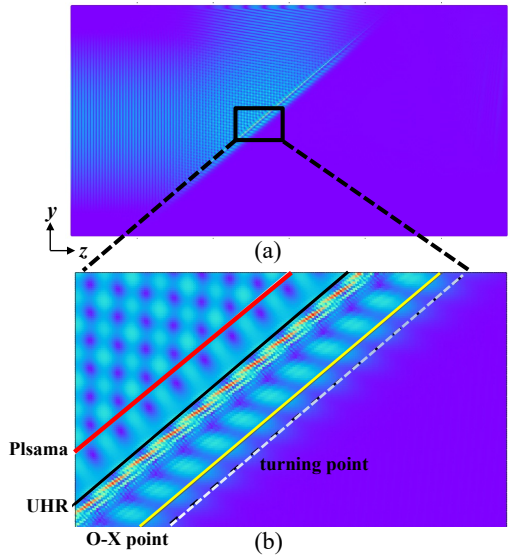

Fig. 5. Electric field intensity in the y–z plane for the optimal incident angle ($\theta = 40.45°$). (a) Global view of the wave propagation showing the O-mode incidence and mode conversion. (b) Enlarged view near the conversion region, indicating the O–X conversion point, the upper hybrid resonance (UHR), and the turning point. The continuous propagation of the wave without significant attenuation is observed.

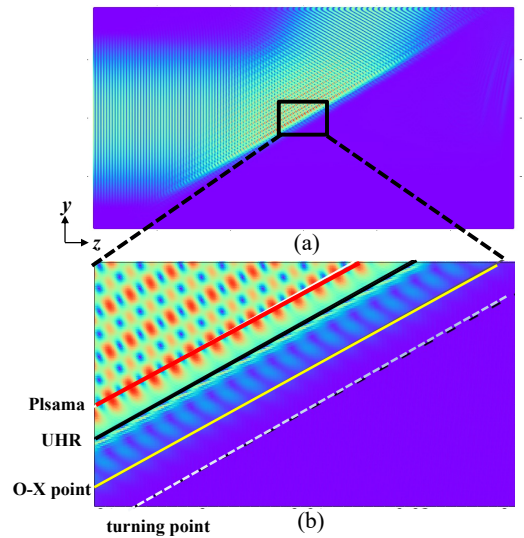

Fig. 6. Electric field intensity in the y–z plane for the non-optimal incident angle ($\theta = 30°$). (a) Global view of the wave propagation showing the O-mode incidence and mode conversion. (b) Enlarged view near the conversion region, indicating the O–X conversion point, the upper hybrid resonance (UHR), and the turning point. In contrast to the optimal case, significant attenuation of the wave is observed due to the presence of an evanescent region.